\documentclass[11pt,a4paper]{article}
\usepackage[hyperref]{emnlp2018}
\usepackage{times}
\usepackage{latexsym}
\usepackage{todonotes}
\usepackage{url}
\usepackage{comment}
\usepackage{adjustbox}
\usepackage{booktabs}
\usepackage{hyperref}
\usepackage{cleveref}
\usepackage{amsmath}
\usepackage{colortbl}

\graphicspath{{graphs/}}

\aclfinalcopy 

\newcommand\blfootnote[1]{%
  \begingroup
  \renewcommand\thefootnote{}\footnote{#1}%
  \addtocounter{footnote}{-1}%
  \endgroup
}
\newcommand{\avsum}{\mathop{\mathpalette\avsuminner\relax}\displaylimits}

\makeatletter
\newcommand\avsuminner[2]{%
  {\sbox0{$\m@th#1\sum$}%
   \vphantom{\usebox0}%
   \ooalign{%
     \hidewidth
     \smash{\vrule height\dimexpr\ht0+1pt\relax depth\dimexpr\dp0+1pt\relax}%
     \hidewidth\cr
     $\m@th#1\sum$\cr
   }%
  }%
}
\makeatother

\title{Belittling the Source: \\ Trustworthiness Indicators to Obfuscate Fake News on the Web}

\author{Diego Esteves \\
        SDA Research \\
        University of Bonn, Germany \\
        {\tt esteves@cs.uni-bonn.de}\\\And
  Aniketh Janardhan Reddy\textsuperscript{+,*}\\
  Carnegie Mellon University\\
  Pittsburgh, USA \\
  {\tt ajreddy@cs.cmu.edu}\\\AND
  Piyush Chawla\textsuperscript{*} \\
  University of Ohio \\
  Ohio, USA \\
  {\tt chawla.81@osu.edu} \\\And
  Jens Lehmann\\
  SDA Research / Fraunhofer IAIS\\
  University of Bonn, Germany \\
  {\tt jens.lehmann@cs.uni-bonn.de}
  }

\date{}

\begin{document}
\maketitle
\begin{abstract}
\blfootnote{\textsuperscript{+}Work was completed while the author was a student at the Birla Institute of Technology and Science, India and was interning at SDA Research.}
\blfootnote{\textsuperscript{*}These two authors contributed equally to this work.}
With the growth of the internet, the number of \textit{fake-news} online has been proliferating every year. The consequences of such phenomena are manifold, ranging from lousy decision-making process to bullying and violence episodes. Therefore, fact-checking algorithms became a valuable asset. To this aim, an important step to detect fake-news is to have access to a credibility score for a given information source. However, most of the widely used Web indicators have either been shut-down to the public (e.g., Google PageRank) or are not free for use (Alexa Rank). Further existing databases are short-manually curated lists of online sources, which do not scale. Finally, most of the research on the topic is theoretical-based or explore confidential data in a restricted simulation environment. In this paper we explore current research, highlight the challenges and propose solutions to tackle the problem of classifying websites into a credibility scale. The proposed model automatically extracts source reputation cues and computes a credibility factor, providing valuable insights which can help in belittling dubious and confirming trustful unknown websites. Experimental results outperform state of the art in the 2-classes and 5-classes setting.
\end{abstract}

\section{Introduction}

With the enormous daily growth of the Web, the number of \textit{fake-news} sources have also been increasing considerably~\cite{Li:2012:TFD:2535568.2448943}. This social network era has provoked a communication revolution that boosted the spread of misinformation, hoaxes, lies and questionable claims. The proliferation of unregulated sources of information allows any person to become an opinion provider with no restrictions. For instance, websites spreading manipulative political content or hoaxes can be persuasive. 
To tackle this problem, different \textit{fact-checking} tools and frameworks have been proposed~\cite{DBLP:journals/corr/ZubiagaABLP17}, mainly divided into two categories: \textit{fact-checking} over natural language claims~\cite{Thorne2018AutomatedFC} and \textit{fact-checking} over knowledge bases, i.e., triple-based approaches~\cite{esteves2018toward}. Overall, \textit{fact-checking} algorithms aim at determining the veracity of claims, which is considered a very challenging task due to the nature of underlying steps, from natural language understanding (e.g. \textit{argumentation mining}) to common-sense verification (i.e., humans have prior knowledge that makes far easier to judge which arguments are plausible and which are not). Yet an important underlying fact-checking step relies upon computing the credibility of sources of information, i.e. indicators that allow answering the question: \textit{``How reliable is a given provider of information}?''. Due to the obvious importance of the Web and the negative impact that misinformation can cause, methods to demote the importance of websites also become a valuable asset. In this sense the high number of new websites appearing at everyday~\cite{statsweb}, make straightforward approaches - such as \textit{blacklists} and \textit{whitelists} - impractical. Moreover, such approaches are not designed to compute credibility scores for a given website but rather to binary label them. Thus, they aim at detecting mostly ``fake'' (threatening) websites; e.g., \textit{phishing detection}, which is out of scope of this work. Thus, open credibility models have a great importance, especially due to the increase of fake news being propagated. There is much research into credibility factors. However, they are mostly grouped as follows: (1) theoretical research on psychological aspects of credibility and (2) experiments performed over private and confidential users information, mostly from web browser activities (strongly supported by private companies). Therefore, while (1) lacks practical results (2) report \textit{findings} which are not much appealing to the broad open-source community, given the non-open characteristic of the conducted experiments and data privacy. Finally, recent research on credibility has also pointed out important drawbacks, as follows:

\begin{enumerate}
    \item Manual (human) annotation of credibility indicators for a set of websites is costly~\cite{haas2017ranking}.
    \item Search engine results page (SERP) do not provide more than few information cues (URL, title and snippet) and the dominant heuristic happens to be the search engine (SE) rank itself~\cite{haas2017ranking}.
    \item Only around 42.67\% of the websites are covered by the credibility evaluation knowledge base, where most domains have a low credibility confidence~\cite{liu2015towards}  
\end{enumerate}

Therefore, automated credibility models play an important role in the community - although not broadly explored yet, in practice. In this paper, we focus on designing computational models to predict the credibility of a given website rather than performing sociological experiments or experiments with end users (simulations). In this scenario, we expect that a website from a domain such as \texttt{bbc.com} gets a higher trustworthiness score compared to one from \texttt{wordpress.com}, for instance.
\label{sec:introduction}

\section{Related Work}

\textit{Credibility} is an important research subject in several different communities and has been the subject of study over the past decades. Most of the research, however, focuses on theoretical aspects of credibility and its persuasive effect on different fundamental problems, such as economic theories~\cite{sobel1985theory}. 
\subsection{Fundamental Research}
\label{subsec:fundamental_research}

A thorough examination of psychological aspects in evaluating documents credibility has been studied~\cite{fogg1999elements,fogg2001makes,fogg2003users}, which reports numerous challenges. Apart from sociological experiments, \textit{Web Credibility} - in a more practical perspective - has a different focus of research, described as follows:\par
\textbf{Rating Systems, Simulations} are mostly platform-based solutions to conduct experiments (mostly using private data) in order to detect credibility factors. Nakamura et al.~\shortcite{Nakamura2007} surveyed internet users from all age groups to understand how they identified trustworthy websites. Based on the results of this survey, they built a graph-based ranking method which helped users in gauging the trustworthiness of search results retrieved by a search engine when issued a query $\mathcal{Q}$. A study by Stanford University revealed important factors that people notice when assessing website credibility~\cite{fogg2003users}, mostly visual aspects (\textit{web site design}, \textit{look} and \textit{information design}). The \textit{writing style} and \textit{bias of information} play a small role as defining the level of credibility (selected by approximately 10\% of the comments). However, this process of evaluating the credibility of web pages by users is impacted only by the number of heuristics they are aware of~\cite{fogg2003prominence}, biasing the human evaluation w.r.t. a limited and specific set features. An important factor considered by humans to judge credibility relies on the search engine results page (SERP). The higher ranked a website is when compared to other retrieved websites the more credible people judge a website to be~\cite{schwarz2011augmenting}. Popularity is yet another major credibility factor~\cite{giudice2010crowdsourcing}.~Liu et al.~\shortcite{liu2015towards} proposed to integrate recommendation functionality into a Web Credibility Evaluation System (WCES), focusing on the user's feedback. 
Shah et al.~\shortcite{shah2015web} propose a full list of important features for credibility aspects, such as 1) the quality of the design of the website and 2) how well the information is structured. In particular, the perceived accuracy of the information was ranked only in 6th place. Thus, superficial website characteristics as heuristics play a key role in credibility evaluation. Dong et al~\shortcite{KBT2015} propose a different method (KBT) to estimate the trustworthiness of a web source based on the information given by the source (i.e., applies fact-checking to infer credibility). This information is represented in the form of triples extracted from the web source. The trustworthiness of the source is determined by the correctness of the triples extracted. Thus, the score is computed based on \textit{endogenous} (e.g., correctness of facts) signals rather then \textit{exogenous} signals (e.g., links). Unfortunately, this research from Google does not provide open data. It is worth mentioning that - surprisingly - their hypothesis (content is more important than visual) contradicts previous research findings~\cite{fogg2003users,shah2015web}. While this might be due to the dynamic characteristic of the Web, this contradiction highlights the need for more research into the real use of web credibility factors w.r.t. automated web credibility models. Similar to~\cite{Nakamura2007}, Singal and Kohli~\shortcite{singal2016trust} proposes a tool (dubbed TNM) to re-rank URLs extracted from Google search engine according to the trust maintained by the actual users). Apart from the search engine API, their tool uses several other APIs to collect website usage information (e.g., traffic and engagement info).~\cite{kakol2017understanding} perform extensive crowdsourcing experiments that contain credibility evaluations, textual comments, and labels for these comments. \par 
\textbf{SPAM/phishing detection}: Abbasi et al.~\shortcite{abbasi2010detecting} propose a set of design guidelines which advocated the development
of SLT-based classification systems for fraudulent website detection, i.e., despite seeming credible - websites that try to obtain private information and defraud visitors. PhishZoo~\cite{phishzoo} is a phishing detection system which helps users in identifying phishing websites which look similar to a given set of protected websites through the creation of profiles.
\subsection{Automated Web Credibility}
\label{subsec:automatedwebcred}
\textbf{Automated Web Credibility} models for website classification are not broadly explored, in practice. The aim is to produce a predictive model given training data (annotated website ranks) regardless of an input query $\mathcal{Q}$. Existing gold standard data is generated from surveys and simulations (see \textit{Rating Systems, Simulations} related work). Currently, state of the art (SOTA) experiments rely on the Microsoft Credibility dataset\footnote{It is worth mentioning that this survey is mostly based on confidential data and it is not available to the open community (e.g., overall popularity, popularity among domain experts, geo-location of users and number of awards)}~\cite{schwarz2011augmenting}. Recent research use the website label (Likert scale) released in the Microsoft dataset as a gold standard to train automated web credibility models, as follows: \par
Olteanu et al.~\shortcite{olteanu2013web} proposes a number of properties (37 linguistic and textual features) and applies machine learning methods to recognize trust levels, obtaining 22 relevant features for the task. Wawer et al.~\shortcite{wawer2014predicting} improve this work using psychosocial and psycholinguistic features (through The General Inquirer (GI) Lexical Database~\cite{stone1966general}) achieving state of the art results. \par
Finally, another resource is the Content Credibility Corpus (C3)~\cite{kakol2017understanding}, the largest Web credibility Corpus publicity available so far. However, in this work authors did not perform experiments w.r.t. \textit{automated credibility models} using a standard measure (i.e., Likert scale), such as in ~\cite{olteanu2013web,wawer2014predicting}. Instead, they rather focused on evaluating the theories of web credibility in order to produce a much larger and richer corpus.

\label{sec:relatedwork}

\section{Experimental Setup}

\subsection{State-of-the-art (SOTA) Features}
\label{subsec:sotafeatures}

Recent research on credibility factors for web sites~\cite{olteanu2013web} have initially divided the features into the following logical groups:

\begin{enumerate}
    \item \textbf{Content-based} (25 features): number of special characters in the text, spelling errors, web site category and etc..
        \begin{enumerate}
            \item \textbf{Text} (20 features)
            \item \textbf{Appearance} (4 features)
            \item \textbf{Meta-information} (1 feature)
        \end{enumerate}
    \item \textbf{Social-based} (12 features): Social Media Metadata (e.g., Facebook shares, Tweets pointing to a certain URL, etc.), Page Rank, Alexa Rank and similar.
        \begin{enumerate}
            \item \textbf{Social Popularity} (9 features)
            \item \textbf{General Popularity} (1 feature)
            \item \textbf{Link structure} (2 features)
        \end{enumerate}
\end{enumerate}

According to~\cite{olteanu2013web}, a resultant number of 22 features (out of 37) were selected as most significant (10 for \textbf{content-based} and all \textbf{social-based} features). Surprisingly (but also following~\cite{KBT2015}), none from the sub-group \textbf{Appearance}, although studies have systematically shown the opposite, i.e., that visual aspects are one of the most important features~\cite{fogg2003users,shah2015web,haas2017ranking}.

In this picture, we claim the most negative aspect is the reliance on \textbf{Social-based} features. This dependency not only affects the final performance of the credibility model, but also implies in financial costs as well as presenting high discriminative capacity, adding a strong bias to the performance of the model\footnote{authors applied ANOVA test confirming this finding}. The computation of these features relies heavily on external (e.g., Facebook API\footnote{\url{https://developers.facebook.com/}} and AdBlock\footnote{\url{https://adblockplus.org/}}) and commercial libraries (Alchemy\footnote{\url{www.alchemyapi.com}}, PageRank\footnote{excepting for heuristic computations, calculation of PageRank requires crawling
the whole Internet}, Alexa Rank\footnote{\url{https://www.alexa.com/siteinfo}}. Thus, engineering and financial costs are a must. Furthermore, popularity on Facebook or Twitter can be measured only by data owners. Additionally, vendors may change the underlying algorithms without further explanation. Therefore, also following Wawer et al.~\shortcite{wawer2014predicting}, in this paper we have excluded \textbf{Social-based} features from our experimental setup.

On top of that,~\cite{wawer2014predicting} incremented the model, adding features extracted from the General Inquirer (GI) Lexical Database, resulting in a vector of 183 extra categories, apart from the selected 22 base features, i.e. total of 205 features (However, this is subject to contradictions. Please see~\Cref{subsec:repro} for more information).

\subsection{Datasets}

\subsubsection{Website credibility evaluation}
\textbf{Microsoft Dataset}~\cite{schwarz2011augmenting} consists of thousands of URLs and their credibility ratings (five-point Likert Scale\footnote{\url{https://en.wikipedia.org/wiki/Likert_scale}}), ranging from 1 ("very non-credible") to 5 ("very credible"). In this study, participants were asked to rate the websites as credible following the definition: ``\textit{A credible webpage is one whose information one can accept as the truth without
needing to look elsewhere}''. Studies by~\cite{olteanu2013web,wawer2014predicting} use this dataset for evaluation.
\textbf{Content Credibility Corpus (C3)\footnote{also known as Reconcile Corpus}} is the most recent and the largest credibility dataset currently publicly available for research~\cite{kakol2017understanding}. It contains 15.750 evaluations of 5.543 URLs from 2.041 participants with some additional information about website characteristics and basic demographic features of users. Among many metadata information existing in the dataset, in this work we are only interested in the URLs and their respective five-point Likert scale, so that we obtain the same information available in the Microsoft dataset.

\subsubsection{Fact-checking influence}

In order to verify the impact of our web credibility model in a real use-case scenario, we ran a fact-checking framework to verify a set of input claims. Then we collected the sources (URLs) containing proofs to support a given claim. We used this as a dataset to evaluate our web credibility model. 

The primary objective is to verify whether our model is able, on average, to assign \textit{lower} scores to the websites that contain \textit{proofs} supporting \textit{claims} which are labeled as \textit{false} in the FactBench dataset (i.e., the source is providing false information, thus should have a lower credibility score). Similarly, we expect that websites that support \textit{positive} claims are assigned with higher scores (i.e., the source is supporting an accurate claim, thus should have a higher credibility score). 

The (gold standard) input claims were obtained from the \textbf{FactBench} dataset\footnote{\url{https://github.com/DeFacto/FactBench}}, a multilingual benchmark for the evaluation of fact validation algorithms. It contains a set of RDF\footnote{\url{https://www.w3.org/RDF/}} models (10 different relations), where each model contains a singular fact expressed as a \textit{subject-predicate-object} triple. The data was automatically extracted from DBpedia and Freebase KBs, and manually curated in order to generate true and false examples.

The website list extraction was carried out by DeFacto~\cite{gerber2015}, a fact-checking framework designed for RDF KBs. DeFacto returns a set of websites as pieces of evidence to support its prediction (true or false) for a given input claim.

\subsection{Final Features}
\label{subsec:finalfeatures}

We implemented a set of \textbf{Content-based} features (\Cref{subsec:sotafeatures}) adding more lexical and textual based features. \textbf{Social-based} features were not considered due to financial costs associated with paid APIs. The final set of features for each website $w$ is defined as follows:\par

1. \textit{Web Archive}: the temporal information w.r.t. cache and freshness. $\Delta_b$ and $\Delta_e$ correspond to the temporal differences of the first and last 2 updates, respectively. $\Delta_a$ represents the age of $w$ and finally $\Delta_u$ represents the temporal difference for the last update to today. $\gamma$ is a penalization factor when the information is obtained from the \textit{domain} of $w$ ($w_d$) instead $w$. 

\[
f_{arc}(w)= \Big( \Big[\frac{1}{\log(\Delta_b \times \Delta_e)} + \log(\Delta_a) + \frac{1}{\Delta_u}\Big] \Big) \times \gamma
\]

2. \textit{Domain}: refers to the (encoded) domain $w$ (e.g. org) \par

3. \textit{Authority}: searches for authoritative keywords within the page HTML content $w_c$ (e.g., contact email, business address, etc..) \par

4. \textit{Outbound Links}: searches the number of different outbound links in $w \wedge w_d \in d$, i.e., $\sum_{n=1}^{P} \phi(w_c)$ where $P$ is the number of web-based protocols. \par

5. \textit{Text Category}: returns a vector containing the probabilities $P$ for each pre-trained category $c$ of $w$ w.r.t. the sentences of the website $w_s$ and page title $w_t$: $\avsum_{s=1}^{w_s} \gamma(s) ^\frown \gamma(w_t)$. We trained a set of binary multinomial Naive Bayes (NB) classifiers, one per class, as follows: \textit{business}, \textit{entertainment}, \textit{politics}, \textit{religion}, \textit{sports} and \textit{tech}.\par

6. \textit{Text Category - LexRank}: reduces the noisy of $w_b$ by classifying only top $N$ sentences generated by applying LexRank~\cite{Erkan:2004:LGL:1622487.1622501} over $w_b$ ($S\prime = \Gamma(w_b, N)$), which is a graph-based text summarizing technique: $\avsum_{s\prime=1}^{S\prime} \gamma(s\prime) ^\frown \gamma(w_t)$. \par

7. \textit{Text Category - LSA}: similarly, we apply Latent Semantic Analysis (LSA)~\cite{Steinberger04usinglatent} to detect semantically important sentences in $w_b$ ($S\prime = \Omega(w_b, N)$): $\avsum_{s\prime=1}^{S\prime} \gamma(s\prime) ^\frown \gamma(w_t)$. \par

8. \textit{Readability Metrics}: returns a vector resulting of the concatenation of several $R$ readability metrics~\cite{Si:2001:SMS:502585.502695}  \par

9. \textit{SPAM}: detects whether the $w_b$ or $w_t$ are classified as spam: $\psi(w_b) ^\frown \psi(w_t)$ \par

10. \textit{Social Tags}: returns the frequency of social tags in $w_b$: $\bigcup\limits_{i=1}^{R} \varphi(i, w_b)$  \par

11. \textit{OpenSources}: returns the open-source classification ($x$) for a given website: 
\[
    x = 
\begin{cases}
    1 ,& \text{if } w \in \mathcal{O} \\
    0 ,& \text{if } w \not\in \mathcal{O}
\end{cases}
\]
\par

12. \textit{PageRankCC}: PageRank information computed through the CommonCrawl\footnote{\url{http://commoncrawl.org/}} Corpus \par

13. \textit{General Inquirer}~\cite{stone1966general}: a 182-lenght vector containing several lexicons\par

14. \textit{Vader Lexicon}: lexicon and rule-based sentiment analysis tool that is specifically attuned to sentiments \par

15. \textit{HTML2Seq}: we introduce the concept of \textit{bag-of-tags}, where similarly to \textit{bag-of-words}\footnote{\url{https://en.wikipedia.org/wiki/Bag-of-words_model}} we group the HTML tag occurrences in each web site. We additionally explore this concept along with a sequence problem, i.e. we encode the tags and evaluate this considering a window size (offset) from the header of the page. \par

\label{sec:features}

\section{Experiments}

Previous research proposes two \textbf{application settings} w.r.t. the classification itself, as follows: (A.1) casting the credibility problem as a classification problem and (A.2) evaluating the credibility on a five-point Likert scale (regression). In the classification scenario, the models are evaluated both w.r.t. the 2-classes as well as 3-classes. In the 2-classes scenario, websites ranging from 1 to 3 are labeled as ``low'' whereas 4 and 5 are labeled as ``high'' (credibility). Analogously, in the 3-classes scenario, websites labeled as 1 and 2 are converted to ``low'', 3 remains as ``medium'' while 4 and 5 are grouped into the ``high'' class. \par
\begin{figure*}[htbp]
  \begin{minipage}[b]{0.5\linewidth}
    \centering
    \includegraphics[width=\linewidth]{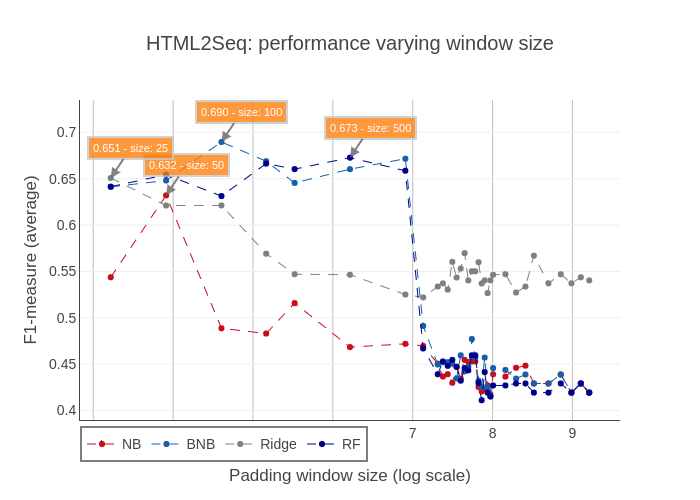}
    \caption{HTML2Seq: F1-measure over different paddings size for the 2-class problem (Microsoft)}
    \label{fig:htmlenc1}
  \end{minipage}
  \hspace{0.5cm}
  \begin{minipage}[b]{0.5\linewidth}
    \centering
    \includegraphics[width=\linewidth]{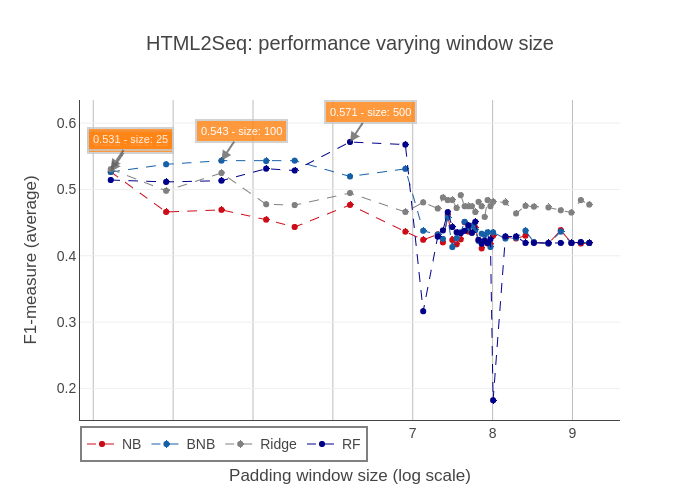}
    \caption{HTML2Seq: F1-measure over different paddings size for the 3-class problem (Microsoft)}
    \label{fig:htmlenc2}
  \end{minipage}
  \begin{minipage}[b]{0.5\linewidth}
    \centering
    \includegraphics[width=\linewidth]{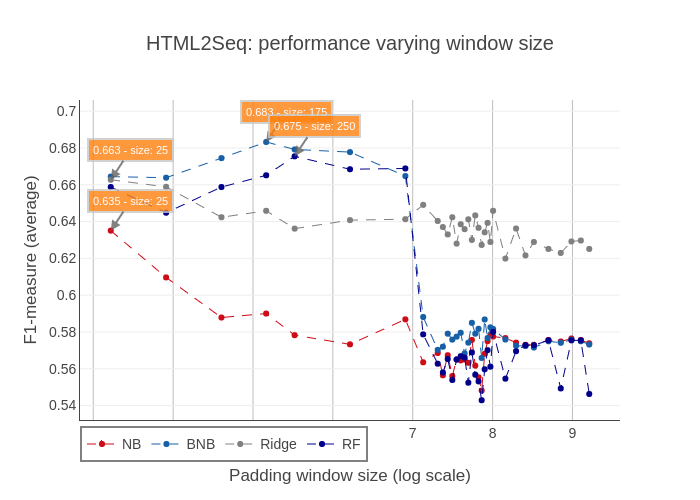}
    \caption{HTML2Seq: F1-measure over different paddings size for the 2-class problem (C3)}
    \label{fig:htmlenc3}
  \end{minipage}
  \hspace{0.5cm}
  \begin{minipage}[b]{0.5\linewidth}
    \centering
    \includegraphics[width=\linewidth]{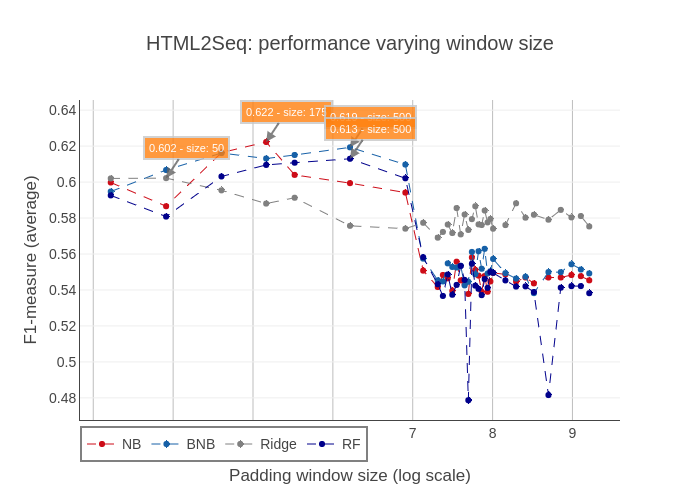}
    \caption{HTML2Seq: F1-measure over different paddings size for the 3-class problem (C3)}
    \label{fig:htmlenc4}
  \end{minipage}
\end{figure*}

We first explore the impact of the \textit{bag-of-tags} strategy. We encode and convert the tags into a sequence of tags, similar to a sequence of sentences (looking for opening and closing tags, e.g., \textless a\textgreater and \textless/a\textgreater). Therefore, we perform document classification over the resulting vectors.~\Cref{fig:htmlenc1,fig:htmlenc2,fig:htmlenc3,fig:htmlenc4} show results of this strategy for both 2 and 3-classes scenarios. The x-axis is the log scale of the paddings (i.e., the offset of HTML tags we retrieved from $w$, ranging from 25 to 10.000). The charts reveal an interesting pattern in both gold-standard datasets (Microsoft Dataset and C3 Corpus): the first tags are the most relevant to predict the credibility class. Although this strategy does not achieve state of the art performance (F1 = 0.690 and 0.571 for the 2 and 3-classes configurations, respectively, when compared to state of the art: F1 = 0.745 and 0.652), it presents reasonable performance by just inspecting website metadata. However, it is worth mentioning that the main advantage of this approach lies in the fact that it is language agnostic (while current research focuses on English) as well as less susceptible to overfitting. \par
\begin{figure*}[htbp]
  \begin{minipage}[b]{0.5\linewidth}
    \centering
    \includegraphics[width=\columnwidth]{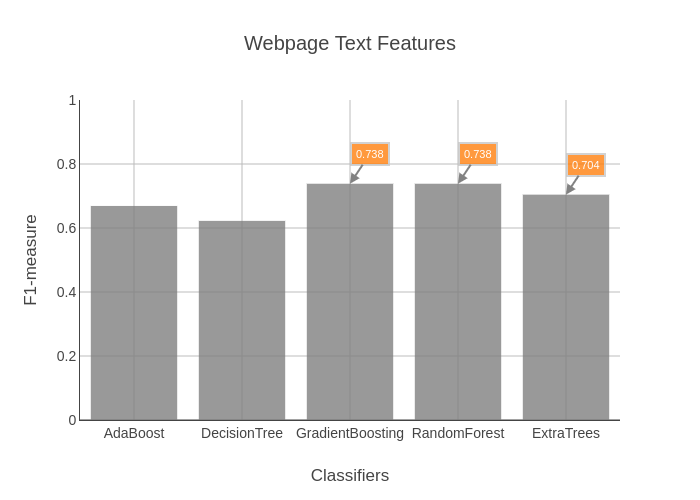}
    \caption{Lexical Features: F1-measure for the 2-class problem (Microsoft). Used \textit{SelectKBest} method for feature selection, K=0.25}
    \label{fig:textfeat}
  \end{minipage}
  \hspace{0.5cm}
  \begin{minipage}[b]{0.5\linewidth}
    \centering
    \includegraphics[width=\columnwidth]{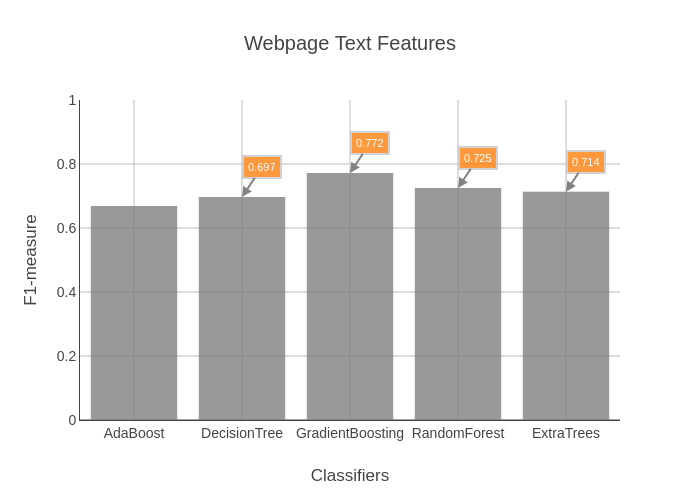}
    \caption{Lexical+HTML2Seq (best padding): F1-measure for the 2-class problem (Microsoft): increasing +3\% (0.772) on average F1 (Gradient Boosting).}
    \label{fig:html2seq_text}
  \end{minipage}
 \end{figure*}

We then evaluate the performance of the textual features (\Cref{subsec:finalfeatures}) isolated. Results for the 2-classes scenario are presented as follows:~\Cref{fig:textfeat} highlights the best models performance using textual features only. While this as a single feature does not outperform the lexical features, when we combine the \textit{bag-of-tags} approach (predictions of probabilities for each class) we boost the performance (F1 from 0.738 to 0.772) and outperform state of the art (0.745), as shown in~\Cref{fig:html2seq_text}.~\Cref{tab1:2class,tab1:3class,tab1:5class} shows detailed results for both datasets (2-classes, 3-classes and 5-classes configurations, respectively). For 5-class regression, we found that the \textit{best pad = 100} for the Microsoft dataset and \textit{best pad = 175} for the C3 Corpus. We preceded the computing of both classification and regression models with feature selection according to a percentile of the highest scoring features (\textit{SelectKBest}). We tested the choice of 3, 5, 10, 25, 50 75 and K=100 percentiles (thus, no selection) of features and did not find a unique K value for every case. It is worth noticing that in general it is easy to detect high credible sources (F1 for ``high'' class around 0.80 in all experiments and both datasets) but recall of ``low'' credible sources is still an issue.

\begin{table}[t]
  \centering
  \begin{tabular}{|l|c|c|c|}
  \toprule
    \multicolumn{4}{|l|}{\textbf{Microsoft Dataset}} \\ 
    \multicolumn{4}{|l|}{(Gradient Boosting, \textit{K} = 25)}\\\midrule
    Class & Precision & Recall & F1 \\\midrule
    low	&0.851 &	0.588&	0.695\\
    high&	0.752	&0.924&	0.829\\\hline
\textit{weighted}	&0.794	&0.781&	0.772\\
\textit{micro}&	0.781&	0.781&	0.781\\
\textit{macro}&	0.801&	0.756&	0.762\\
    \midrule
    \multicolumn{4}{|l|}{\textbf{C3 Corpus}}\\
    \multicolumn{4}{|l|}{(AdaBoost, \textit{K} = 75)}\\
    \midrule
    Class & Precision & Recall & F1 \\\midrule
    low & 0.558	& 0.355	& 0.434 \\
    high & 0.732 & 0.862 & 0.792 \\\hline
    \textit{weighted} &	0.675 &	0.695 &	0.674 \\
    \textit{micro} & 0.695 & 0.695 & 0.695 \\
    \textit{macro} & 0.645 & 0.609 & 0.613 \\
    \bottomrule
  \end{tabular}
  \caption{Text+HTML2Seq features (2-class): best classifier performance}
  \label{tab1:2class}
\end{table}

\begin{table}[t]
  \centering
  \begin{tabular}{|l|c|c|c|}
  \toprule
    \multicolumn{4}{|l|}{\textbf{Microsoft Dataset}} \\
    \multicolumn{4}{|l|}{(Gradient Boosting, \textit{K} = 75)}\\
    \midrule
    Class & Precision & Recall & F1 \\\midrule
     low	& 0.567	& 0.447	& 0.500 \\
medium& 	0.467	& 0.237& 	0.315 \\
high& 	0.714	& 0.916& 	0.803 \\\hline
\textit{weighted}& 	0.626	& 0.662	& 0.626 \\
\textit{micro}	& 0.662	& 0.662	& 0.662 \\
\textit{macro}	& 0.583	& 0.534	& 0.539 \\
    \midrule
    \multicolumn{4}{|l|}{\textbf{C3 Corpus}} \\\multicolumn{4}{|l|}{(AdaBoost, \textit{K} = 100)}\\\midrule
    Class & Precision & Recall & F1 \\\midrule
         low	&0.143	&0.031&	0.051 \\
medium	&0.410	&0.177	&0.247\\
high	&0.701&	0.916&	0.794\\\hline
\textit{weighted}&	0.583&	0.660	&0.598\\
\textit{micro}&	0.660&	0.660&	0.660\\
\textit{macro}&	0.418&	0.375&	0.364\\
    \bottomrule
  \end{tabular}
  \caption{Text+HTML2Seq features (3-class): best classifier performance}
  \label{tab1:3class}
\end{table}

\begin{table}[t]
  \centering
  \begin{tabular}{|l|c|c|c|c|c|}
  \toprule
    \multicolumn{6}{|l|}{\textbf{Microsoft Dataset}} \\\midrule
    model & $K$ & $R^2$ & RMSE & MAE & EVar\\\midrule
    SVR & 3 & 0.232	& 0.861&	0.691 & 0.238 \\
    Ridge & 3 & 0.268 &	0.841 &	0.683 & 0.269 \\
    \midrule
    \multicolumn{6}{|l|}{\textbf{C3 Corpus}} \\\midrule
    model & $K$ & $R^2$ & RMSE & MAE & EVar \\\midrule
  SVR & 25 & 0.096	& 0.939	& 0.739 & 0.102  \\
  Ridge & 25 & 0.133&	0.920&	0.750 & 0.134 \\
    \bottomrule
  \end{tabular}
  \caption{Text+HTML2Seq: regression measures (5-class). Selecting top $K$ lexical features}
  \label{tab1:5class}
\end{table}

\Cref{tab5} shows statistics on the data generated by the fact-checking algorithm. For 1500 claims, it collected pieces of evidence for over 27.000 websites.~\Cref{tab6} depicts the impact of the credibility model in the fact-checking context. We collected a small subset of 186 URLs from the FactBench dataset and manually annotated\footnote{By four human annotators. In the event of a tie we exclude the URL from the final dataset.} the credibility for each URL (following the Likert scale). The model corrected labeled around 80\% of the URLs associated with a positive claim and, more importantly, 70\% of non-credible websites linked to false claims were correctly identified. This helps to minimize the number of non-credible information providers that contain information that supports a \textit{false} claim.

\begin{table}[t]
  \centering
  \begin{tabular}{|l|c|c|c|}
  \toprule
    \multicolumn{3}{|l|}{FactBench (Credibility Model)} \\\midrule
    \textbf{label} & \textbf{claims} & \textbf{sites} \\\midrule
     true & 750 & 14.638  \\ 
     false & 750 & 13.186  \\\hline
     - & 1500 & 27.824  \\
    \bottomrule
  \end{tabular}
  \caption{FactBench: Web sites collected from claims}
  \label{tab5}
\end{table}

\begin{table}[t]
  \centering
  \begin{tabular}{|l|c|c|c|c|}
  \toprule
    \multicolumn{5}{|l|}{FactBench (Sample - Human Annotation)} \\\midrule
    \textbf{label} & \textbf{claims} & \textbf{sites} & \textbf{non-cred} & \textbf{cred}  \\\midrule
     true & 5 & 96 & \cellcolor{gray!20}57 & 39  \\ 
     false & 5 & 80 & \cellcolor{gray!20}48 & 32  \\\hline
     - & 10 & 186 & \cellcolor{gray!20}105 & 71 \\
     \midrule
     \multicolumn{5}{|l|}{FactBench (Sample - Credibility Model)} \\\midrule
    \textbf{label} & \textbf{non-cred} & \textbf{\%} & \textbf{cred} & \textbf{\%}  \\\midrule
     true & \cellcolor{gray!20}40 & \cellcolor{gray!20}0.81 & 31 & 0.79  \\ 
     false & \cellcolor{gray!20}34 & \cellcolor{gray!20}0.70 & 24 & 0.75  \\
    \bottomrule
  \end{tabular}
  \caption{FactBench Dataset: analyzing the performance of the credibility model in the fact-checking task}
  \label{tab6}
\end{table}

\subsection{Discussion}
\label{subsec:repro}

Reproducibility is still one of the cornerstones of science and scientific projects~\cite{baker20161}. In the following, we list some relevant issues encountered while performing our experiments: \par 

\textbf{Experimental results}: this gap is also observed w.r.t. results reported by~\cite{olteanu2013web}, which is acknowledged by~\cite{wawer2014predicting}, despite numerous attempts to replicate experiments. Authors~\cite{wawer2014predicting} believe this is due to the lack of parameters and hyperparameters explicitly cited in the previous research~\cite{olteanu2013web}. \par 

\textbf{Microsoft dataset}: presents inconsistencies. Although all the web pages are cached (in theory) in order to guarantee a deterministic environment, the dataset - in its original form\footnote{The original dataset can be downloaded from \url{http://research.microsoft.com/en-us/projects/credibility/}} - has a number of problems, as follows: (a) web pages not physically cached (b) URL not matching (dataset links \textit{versus} cached files) (c) Invalid file format (e.g., PDF). Even though these issues have also been previously identified by related research~\cite{olteanu2013web} it is not clear what the URLs for the final dataset (i.e., the support) are nor where this new version is available. \par 

\textbf{Contradictions}: w.r.t. the divergence of the importance of visual features have drawn our attention~\cite{KBT2015} and ~\cite{fogg2003prominence,shah2015web} which corroborate to the need of more methods to solve the web credibility problem, in practice. The main hypothesis that supports this contradiction relies on the fact that feature-based credibility evaluation eventually ignites cat-and-mouse play between scientists and people interested in manipulating the models. In this case, \textit{reinforcement learning} methods pose as a good alternative for adaptation.

\textbf{Proposed features}: The acknowledgement made by authors in~\cite{wawer2014predicting} that ``\textit{solutions based purely on external APIs are difficult to use beyond scientific application and are prone for manipulation}'' confirming the need to exclude \textbf{social features} from research of~\cite{olteanu2013web} contradicts itself. In the course of experiments, authors admit the usage of all features proposed by~\cite{olteanu2013web}: ``\textit{Table 1 presents regression results for the dataset described in [13] in its original version (37 features) and extended with 183 variables from the General Inquirer (to 221 features)}''. 

Therefore, due to the number of relevant issues presented w.r.t. reproducibility and contradiction of arguments, the comparison to recent research becomes more difficult. In this work, we solved the technical issues in the Microsoft dataset and released a new fixed version\footnote{more information at the project website: \url{https://github.com/DeFacto/WebCredibility}}. Also, since we need to perform evaluations in a deterministic environment, we cached and released the websites for the C3 corpus. After scraping, 2.977 URLs were used (out of 5.543). Others were left due to processing errors (e.g., 404). The algorithms and its hyperparameters and further relevant metadata are available through the MEX Interchange Format~\cite{Esteves:2015:MVL:2814864.2814883}. By doing this, we provide a computational environment to perform safer comparisons, being engaged in recent discussions about mechanisms to measure and enhance the reproducibility of scientific projects~\cite{wilkinson2016fair}.
\label{sec:experiments}

\section{Conclusion}

In this work, we discuss existing alternatives, gaps and current challenges to tackle the problem of web credibility. More specifically, we focused on automated models to compute a credibility factor for a given website. This research follows the former studies presented by~\cite{olteanu2013web,wawer2014predicting} and presents several contributions. First, we propose different features to avoid the financial cost imposed by external APIs in order to access website credibility indicators. This issue has become even more relevant in the light of the challenges that have emerged after the shutdown of Google PageRank, for instance. To bridge this gap, we have proposed the concept of bag-of-tags. Similar to~\cite{wawer2014predicting}, we conduct experiments in a highly-dimensional feature space, but also considering web page metadata, which outperforms state of the art results in the 2-classes and 5-classes settings. Second, we identified and fixed several problems on a gold standard dataset for web credibility (Microsoft), as well as indexed several web pages for the C3 Corpus.
Finally, we evaluate the impact of the model in a real fact-checking use-case. We show that the proposed model can help in belittling and supporting different websites that contain evidence of true and false claims, which helps the very challenging fact verification task. As future work, we plan to explore deep learning methods over the HTML2Seq module.

\label{sec:conclusion}

\section*{Acknowledgments}
This research was partially supported by an EU H2020 grant provided for the WDAqua project (GA no. 642795) and by DAAD under the project ``International promovieren in Deutschland – für alle'' (IPID4all).

\bibliography{web-bibliography}

\begin{thebibliography}{31}
\expandafter\ifx\csname natexlab\endcsname\relax\def\natexlab#1{#1}\fi

\bibitem[{Abbasi et~al.(2010)Abbasi, Zhang, Zimbra, Chen, and
  Nunamaker~Jr}]{abbasi2010detecting}
Ahmed Abbasi, Zhu Zhang, David Zimbra, Hsinchun Chen, and Jay~F Nunamaker~Jr.
  2010.
\newblock Detecting fake websites: the contribution of statistical learning
  theory.
\newblock \emph{Mis Quarterly}, pages 435--461.

\bibitem[{Afroz and Greenstadt(2011)}]{phishzoo}
Sadia Afroz and Rachel Greenstadt. 2011.
\newblock Phishzoo: Detecting phishing websites by looking at them.
\newblock \emph{2012 IEEE Sixth International Conference on Semantic
  Computing}, 00:368--375.

\bibitem[{Baker(2016)}]{baker20161}
Monya Baker. 2016.
\newblock 1,500 scientists lift the lid on reproducibility.
\newblock \emph{Nature News}, 533(7604):452.

\bibitem[{Dong et~al.(2015)Dong, Gabrilovich, Murphy, Dang, Horn, Lugaresi,
  Sun, and Zhang}]{KBT2015}
Xin~Luna Dong, Evgeniy Gabrilovich, Kevin Murphy, Van Dang, Wilko Horn, Camillo
  Lugaresi, Shaohua Sun, and Wei Zhang. 2015.
\newblock Knowledge-based trust: Estimating the trustworthiness of web sources.
\newblock \emph{Proc. VLDB Endow.}, 8(9):938--949.

\bibitem[{Erkan and Radev(2004)}]{Erkan:2004:LGL:1622487.1622501}
G\"{u}nes Erkan and Dragomir~R. Radev. 2004.
\newblock Lexrank: Graph-based lexical centrality as salience in text
  summarization.
\newblock \emph{J. Artif. Int. Res.}, 22(1):457--479.

\bibitem[{Esteves et~al.(2015)Esteves, Moussallem, Neto, Soru, Usbeck,
  Ackermann, and Lehmann}]{Esteves:2015:MVL:2814864.2814883}
Diego Esteves, Diego Moussallem, Ciro~Baron Neto, Tommaso Soru, Ricardo Usbeck,
  Markus Ackermann, and Jens Lehmann. 2015.
\newblock Mex vocabulary: A lightweight interchange format for machine learning
  experiments.
\newblock In \emph{Proceedings of the 11th International Conference on Semantic
  Systems}, SEMANTICS '15, pages 169--176, New York, NY, USA. ACM.

\bibitem[{Esteves et~al.(2018)Esteves, Rula, Reddy, and
  Lehmann}]{esteves2018toward}
Diego Esteves, Anisa Rula, Aniketh~Janardhan Reddy, and Jens Lehmann. 2018.
\newblock Toward veracity assessment in rdf knowledge bases: An exploratory
  analysis.
\newblock \emph{Journal of Data and Information Quality (JDIQ)}, 9(3):16.

\bibitem[{Fogg et~al.(2001)Fogg, Marshall, Laraki, Osipovich, Varma, Fang,
  Paul, Rangnekar, Shon, Swani et~al.}]{fogg2001makes}
BJ~Fogg, Jonathan Marshall, Othman Laraki, Alex Osipovich, Chris Varma,
  Nicholas Fang, Jyoti Paul, Akshay Rangnekar, John Shon, Preeti Swani, et~al.
  2001.
\newblock What makes web sites credible?: a report on a large quantitative
  study.
\newblock In \emph{Proceedings of the SIGCHI conference on Human factors in
  computing systems}, pages 61--68. ACM.

\bibitem[{Fogg and Tseng(1999)}]{fogg1999elements}
BJ~Fogg and Hsiang Tseng. 1999.
\newblock The elements of computer credibility.
\newblock In \emph{Proceedings of the SIGCHI conference on Human Factors in
  Computing Systems}, pages 80--87. ACM.

\bibitem[{Fogg(2003)}]{fogg2003prominence}
Brian~J Fogg. 2003.
\newblock Prominence-interpretation theory: Explaining how people assess
  credibility online.
\newblock In \emph{CHI'03 extended abstracts on human factors in computing
  systems}, pages 722--723. ACM.

\bibitem[{Fogg et~al.(2003)Fogg, Soohoo, Danielson, Marable, Stanford, and
  Tauber}]{fogg2003users}
Brian~J Fogg, Cathy Soohoo, David~R Danielson, Leslie Marable, Julianne
  Stanford, and Ellen~R Tauber. 2003.
\newblock How do users evaluate the credibility of web sites?: a study with
  over 2,500 participants.
\newblock In \emph{Proceedings of the 2003 conference on Designing for user
  experiences}, pages 1--15. ACM.

\bibitem[{Gerber et~al.(2015)Gerber, Esteves, Lehmann, B{\"u}hmann, Usbeck,
  {Ngonga Ngomo}, and Speck}]{gerber2015}
Daniel Gerber, Diego Esteves, Jens Lehmann, Lorenz B{\"u}hmann, Ricardo Usbeck,
  Axel-Cyrille {Ngonga Ngomo}, and Ren{\'e} Speck. 2015.
\newblock Defacto - temporal and multilingual deep fact validation.
\newblock \emph{Web Semantics: Science, Services and Agents on the World Wide
  Web}.

\bibitem[{Giudice(2010)}]{giudice2010crowdsourcing}
Katherine~Del Giudice. 2010.
\newblock Crowdsourcing credibility: The impact of audience feedback on web
  page credibility.
\newblock In \emph{Proceedings of the 73rd ASIS\&T Annual Meeting on Navigating
  Streams in an Information Ecosystem-Volume 47}, page~59. American Society for
  Information Science.

\bibitem[{Haas and Unkel(2017)}]{haas2017ranking}
Alexander Haas and Julian Unkel. 2017.
\newblock Ranking versus reputation: perception and effects of search result
  credibility.
\newblock \emph{Behaviour \& Information Technology}, 36(12):1285--1298.

\bibitem[{Kakol et~al.(2017)Kakol, Nielek, and
  Wierzbicki}]{kakol2017understanding}
Michal Kakol, Radoslaw Nielek, and Adam Wierzbicki. 2017.
\newblock Understanding and predicting web content credibility using the
  content credibility corpus.
\newblock \emph{Information Processing \& Management}, 53(5):1043--1061.

\bibitem[{Li et~al.(2012)Li, Dong, Lyons, Meng, and
  Srivastava}]{Li:2012:TFD:2535568.2448943}
Xian Li, Xin~Luna Dong, Kenneth Lyons, Weiyi Meng, and Divesh Srivastava. 2012.
\newblock Truth finding on the deep web: Is the problem solved?
\newblock \emph{Proc. VLDB Endow.}, 6(2):97--108.

\bibitem[{Liu et~al.(2015)Liu, Nielek, Adamska, Wierzbicki, and
  Aberer}]{liu2015towards}
Xin Liu, Radoslaw Nielek, Paulina Adamska, Adam Wierzbicki, and Karl Aberer.
  2015.
\newblock Towards a highly effective and robust web credibility evaluation
  system.
\newblock \emph{Decision Support Systems}, 79:99--108.

\bibitem[{Nakamura et~al.(2007)Nakamura, Konishi, Jatowt, Ohshima, Kondo,
  Tezuka, Oyama, and Tanaka}]{Nakamura2007}
Satoshi Nakamura, Shinji Konishi, Adam Jatowt, Hiroaki Ohshima, Hiroyuki Kondo,
  Taro Tezuka, Satoshi Oyama, and Katsumi Tanaka. 2007.
\newblock \emph{Trustworthiness Analysis of Web Search Results}. Springer
  Berlin Heidelberg, Berlin, Heidelberg.

\bibitem[{Netcraft(2016)}]{statsweb}
Netcraft. 2016.
\newblock Netcraft survey (2016).
\newblock \url{http://www.webcitation.org/6lhJlHtez}.
\newblock Accessed: 2017-10-01.

\bibitem[{Olteanu et~al.(2013)Olteanu, Peshterliev, Liu, and
  Aberer}]{olteanu2013web}
Alexandra Olteanu, Stanislav Peshterliev, Xin Liu, and Karl Aberer. 2013.
\newblock Web credibility: features exploration and credibility prediction.
\newblock In \emph{European conference on information retrieval}, pages
  557--568. Springer.

\bibitem[{Schwarz and Morris(2011)}]{schwarz2011augmenting}
Julia Schwarz and Meredith Morris. 2011.
\newblock Augmenting web pages and search results to support credibility
  assessment.
\newblock In \emph{Proceedings of the SIGCHI Conference on Human Factors in
  Computing Systems}, pages 1245--1254. ACM.

\bibitem[{Shah et~al.(2015)Shah, Ravana, Hamid, and Ismail}]{shah2015web}
Asad~Ali Shah, Sri~Devi Ravana, Suraya Hamid, and Maizatul~Akmar Ismail. 2015.
\newblock Web credibility assessment: affecting factors and assessment
  techniques.
\newblock \emph{Information Research}, 20(1):20--1.

\bibitem[{Si and Callan(2001)}]{Si:2001:SMS:502585.502695}
Luo Si and Jamie Callan. 2001.
\newblock A statistical model for scientific readability.
\newblock In \emph{Proceedings of the Tenth International Conference on
  Information and Knowledge Management}, CIKM '01, pages 574--576, New York,
  NY, USA. ACM.

\bibitem[{Singal and Kohli(2016)}]{singal2016trust}
Himani Singal and Shruti Kohli. 2016.
\newblock Trust necessitated through metrics: Estimating the trustworthiness of
  websites.
\newblock \emph{Procedia Computer Science}, 85:133--140.

\bibitem[{Sobel(1985)}]{sobel1985theory}
Joel Sobel. 1985.
\newblock A theory of credibility.
\newblock \emph{The Review of Economic Studies}, 52(4):557--573.

\bibitem[{Steinberger and Ježek(2004)}]{Steinberger04usinglatent}
Josef Steinberger and Karel Ježek. 2004.
\newblock Using latent semantic analysis in text summarization and summary
  evaluation.
\newblock In \emph{In Proc. ISIM ’04}, pages 93--100.

\bibitem[{Stone and Hunt(1963)}]{stone1966general}
Philip~J. Stone and Earl~B. Hunt. 1963.
\newblock A computer approach to content analysis: Studies using the general
  inquirer system.
\newblock In \emph{Proceedings of the May 21-23, 1963, Spring Joint Computer
  Conference}, AFIPS '63 (Spring), pages 241--256, New York, NY, USA. ACM.

\bibitem[{Thorne and Vlachos(2018)}]{Thorne2018AutomatedFC}
James Thorne and Andreas Vlachos. 2018.
\newblock Automated fact checking: Task formulations, methods and future
  directions.
\newblock \emph{CoRR}, abs/1806.07687.

\bibitem[{Wawer et~al.(2014)Wawer, Nielek, and
  Wierzbicki}]{wawer2014predicting}
Aleksander Wawer, Radoslaw Nielek, and Adam Wierzbicki. 2014.
\newblock Predicting webpage credibility using linguistic features.
\newblock In \emph{Proceedings of the 23rd international conference on world
  wide web}, pages 1135--1140. ACM.

\bibitem[{Wilkinson et~al.(2016)Wilkinson, Dumontier, Aalbersberg, Appleton,
  Axton, Baak, Blomberg, Boiten, da~Silva~Santos, Bourne
  et~al.}]{wilkinson2016fair}
Mark~D Wilkinson, Michel Dumontier, IJsbrand~Jan Aalbersberg, Gabrielle
  Appleton, Myles Axton, Arie Baak, Niklas Blomberg, Jan-Willem Boiten,
  Luiz~Bonino da~Silva~Santos, Philip~E Bourne, et~al. 2016.
\newblock The fair guiding principles for scientific data management and
  stewardship.
\newblock \emph{Scientific data}, 3.

\bibitem[{Zubiaga et~al.(2017)Zubiaga, Aker, Bontcheva, Liakata, and
  Procter}]{DBLP:journals/corr/ZubiagaABLP17}
Arkaitz Zubiaga, Ahmet Aker, Kalina Bontcheva, Maria Liakata, and Rob Procter.
  2017.
\newblock Detection and resolution of rumours in social media: {A} survey.
\newblock \emph{CoRR}, abs/1704.00656.

\end{thebibliography}
\bibliographystyle{acl_natbib_nourl}

\end{document}